\documentclass{kapproc} 

\usepackage{procps} 

\usepackage[dvips]{graphicx}

\upperandlowercase

\kluwerbib 

\newcommand\etal{et~al.}
\newcommand\gta{\mathrel{\hbox{\rlap{\hbox{\lower3pt\hbox{$\sim$}}}\raise1pt\hbox{$>$}}}}
\newcommand\gtrsim{\gta}

\begin{document}

\articletitle[The Evolutionary Status of Clusters of Galaxies at z $\sim 1$]
{The Evolutionary Status of Clusters of Galaxies at z $\sim$ 1}

\author{Holland Ford\altaffilmark{1}, Marc Postman\altaffilmark{2},
J.~P.~Blakeslee\altaffilmark{1}, R.~Demarco\altaffilmark{1},
M.~J.~Jee\altaffilmark{1}, P.~Rosati\altaffilmark{4},
B.~P.~Holden\altaffilmark{3}, N.~Homeier\altaffilmark{1},  
G.~Illingworth\altaffilmark{3}, R.~L.~White\altaffilmark{2} }
 
\affil{\altaffilmark{1}Johns Hopkins University, Baltimore, MD, USA  
\altaffilmark{2}Space Telescope Science Institute, Baltimore, MD, USA
\altaffilmark{3}Lick Observatory, University of California, Santa Cruz, CA 95064
\altaffilmark{4}Karl-Schwarzschild-Strasse 2, D-85748 Garching, Germany}

\begin{abstract}
Combined HST, X-ray, and ground-based optical studies show that
clusters of galaxies are largely "in place" by $z \sim 1$, an epoch
when the Universe was less than half its present age.  High resolution
images show that elliptical, S0, and spiral galaxies are present in
clusters at redshifts up to $z \sim 1.3$. Analysis of the CMDs 
suggest that the cluster ellipticals formed
their stars several Gyr earlier, at $z \gtrsim 3$.
The morphology--density relation is well established at $z \sim 1$, with 
star-forming spirals and irregulars residing mostly in the outer parts of
the clusters and E/S0s
concentrated in dense clumps.  The intra\-cluster medium has 
already reached the metallicity of present-day clusters.
The distributions of the hot gas and early-type galaxies are similar in $z \sim 1$ 
clusters, indicating both have largely virialized in the deepest potentials wells.

In spite of the many similarities between $z \sim 1$ and present-day
clusters, there are significant differences.  The morphologies
revealed by the hot gas, and particularly the early-type galaxies,
are elongated rather than spherical.   We appear to be
observing the clusters at an epoch when the sub-clusters and groups
are still assembling into a single regular cluster.  Support for this
picture comes from CL0152 where the gas appears to be lagging behind
the luminous and dark mass in two merging sub-components. 
Moreover, the luminosity difference between the first and second 
brightest cluster galaxies at $z\sim1$ is smaller than in 93\% of present-day 
Abell clusters, which suggests that considerable luminosity evolution
through merging has occurred since that epoch.
Evolution is also seen in the bolometric X-ray luminosity function.
\end{abstract}


\section{Introduction}

The Advanced Camera for Surveys (ACS) IDT is using the new
capabilities of the ACS (Ford \etal\ 2002) to answer fundamental
questions about clusters and cluster galaxies at redshifts $z \sim 1$,
an epoch when they are approximately half the age of the Universe.
Our goals include constraining the formation ages and the SF history
of early-type galaxies, measuring the fundamental properties of
cluster galaxies (e.g. structure, morphology, and luminosity) and
their relationships within the clusters, measuring the evolution of
cluster and galaxy characteristics from $z \sim 1$ to the present, and
investigating the assembly of the brightest cluster galaxies.  We also
aim to establish links between clusters at $z \sim 1$ and
proto-clusters at $z \sim 2$ to 5 (Miley \etal\ 2004),
though this will not be discussed here. In this paper we describe the
results to date from our ongoing study of eight clusters at $z \sim
1$.  In section 3 we discuss CL1252 and CL0152 in detail, the two
clusters where our analysis has progressed furthest. In subsequent
sections we discuss and compare properties of the entire sample, and
then end with a discussion of the evolutionary status of the clusters.

\section
{Cluster Selection and Cluster Properties}

Five of the clusters in our program were initially identified from the
ROSAT Deep Cluster Survey (Rosati \etal\ 1998; RDCS), one from the
Einstein Extended Medium Sensitivity Survey (Gioia \& Luppino 1994;
MS1054), and two from a Palomar deep near-infrared photographic survey
(CL1604+4304 \& CL1604 +4321; Gunn \etal\ 1986). The reality of the
clusters has been confirmed by extensive spectroscopy with ground
based telescopes.  The properties of the clusters and their ACS
observations are summarized in Table \ref{cluster_properties}.  The
number of spectroscopically confirmed galaxies in each cluster is in
parentheses in column 2.  The velocity dispersions in column 3 are in
the clusters' rest frames.  The age of the Universe $T_z$ at redshift
$z$ assumes $h = 0.7$, $\Omega_m = 0.30$, $\Omega_\Lambda = 0.70$,
giving $T_0{\,=\,13.47}$ Gyr today, the cosmology we use throughout
this paper unless stated otherwise.

\begin{table}[ht]
\caption[Properties of High Redshift Clusters Observed with ACS.]
{Properties of High Redshift Clusters Observed with ACS.}
\begin{tabular*}{\textwidth}{@{\extracolsep{\fill}}lccccc}
\sphline
\it Cluster &\it Redshift$^a$ &\it Rest Frame &\it X-ray Lum.  &\it Filters$^b$ &\it Total\cr
\it  &\it $T_z$(Gyr) &\it Vel. Dispersion&\it  ($10^{44}$ ergs s$^{-1}$) &\it &\it Orbits\cr
\sphline
MS1054&	0.831(143)~6.5&	1112&	23.3&	V,i,z&	24\cr
CL0152&	0.837 (~102)~6.5&	1632$^c$&	7.8&	r,i,z&	24\cr
CL1604+4304&	0.897 (~22)~6.2&	1226&	2.0&	V,I&	4\cr
CL1604+4321&	0.924 (~44)~6.1&	935&	<1.2&	V,I&	4\cr
CL0910&	1.101 (~10)~5.4&	N/A&	1.5&	i,z&	8\cr
CL1252&	1.237 (~36)~4.9&	760$^d$&	2.5&	i,z&	32\cr
CL0848-A,B&	1.265 (~40)~4.8&	640 (A)&	1.5 (A) {}{}   $\sim$1 (B)&	i,z&	24\cr
\sphline
\label{cluster_properties}
\end{tabular*}
\begin{tablenotes}
$^a$The number of spectroscopically confirmed members is in parentheses.\\
$^b$Capital letters are Johnson Filters and small case letters are Sloan filters.\\
$^c$Demarco \etal\ 2004\\
$^d$Girardi \etal\ 2004
\end{tablenotes}
\vspace{-1cm}
\end{table}

\section
{The Clusters CL1252 and CL0152}
\subsection
{CL1252}

\begin{figure}[ht]
\centerline{\includegraphics[width=4.5in]{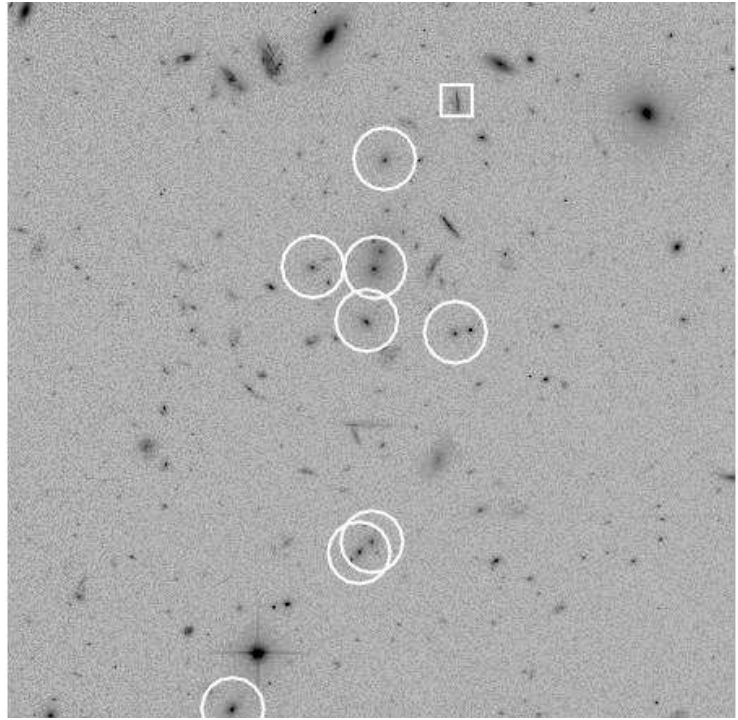}}
\caption{An ACS composite i,z image of the center of the cluster
CL1252 at $z = 1.237$.  The field is 70$^{\prime\prime}$ square ($\sim
580$ kpc in the restframe).  Spectroscopically confirmed members are
circled; emission line galaxies are boxed.  The circles are
6$^{\prime\prime}$ (50 kpc) in diameter. The "red" early-type cluster
members are very conspicuous in a composite ACS i,z/VLT-K image.}
\label{cl1252_ctr}
\end{figure}

\begin{figure}[ht]
\centerline{\includegraphics[width=4.5in]{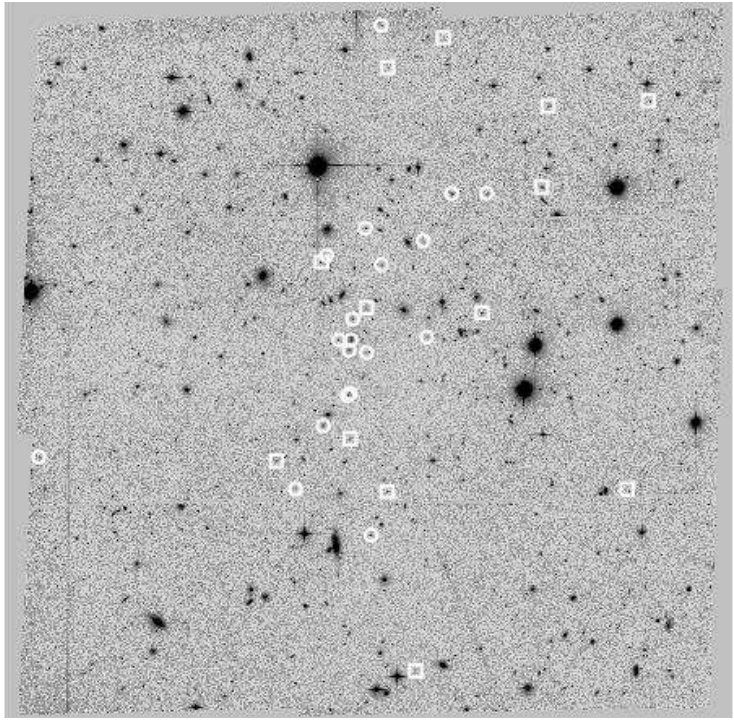}}
\caption{An ACS composite i,z mosaic image of four overlapping fields
centered on CL1252 (symbols as in Figure 1).  
The field is $\sim 350^{\prime\prime}$ on a side
($\sim 2.9$ Mpc in the restframe).}
\label{cl1252_all}
\end{figure}

Figure \ref{cl1252_ctr} shows the center of one of our most distant
clusters, CL1252 at $z = 1.237$.  Spectroscopically confirmed members
are marked with circles (passive galaxies) and squares (emission line
galaxies).  Figure \ref{cl1252_all} shows the confirmed cluster
members in a combined $i,z$ mosaic image of four overlapping ACS
pointings.  The two figures show several facts. The cluster is large;
the most distant confirmed members are $\sim 170^{\prime\prime}$
($\sim 1.4$ Mpc) from the "center" of the cluster.  Like other
clusters at this redshift, the projected image of CL1252 is elongated.
Finally, the passive, primarily early-type galaxies, are more strongly
concentrated to the center and to the axis of the cluster than are the
later type galaxies with [OII]$\lambda$3727 emission, i.e. the
star-forming galaxies.  The X-ray and lensing results discussed below
show that the early-type galaxies are primarily in
the deepest part of the cluster's potential.

\begin{figure}[ht]
\centerline{\includegraphics[width=4.5in]{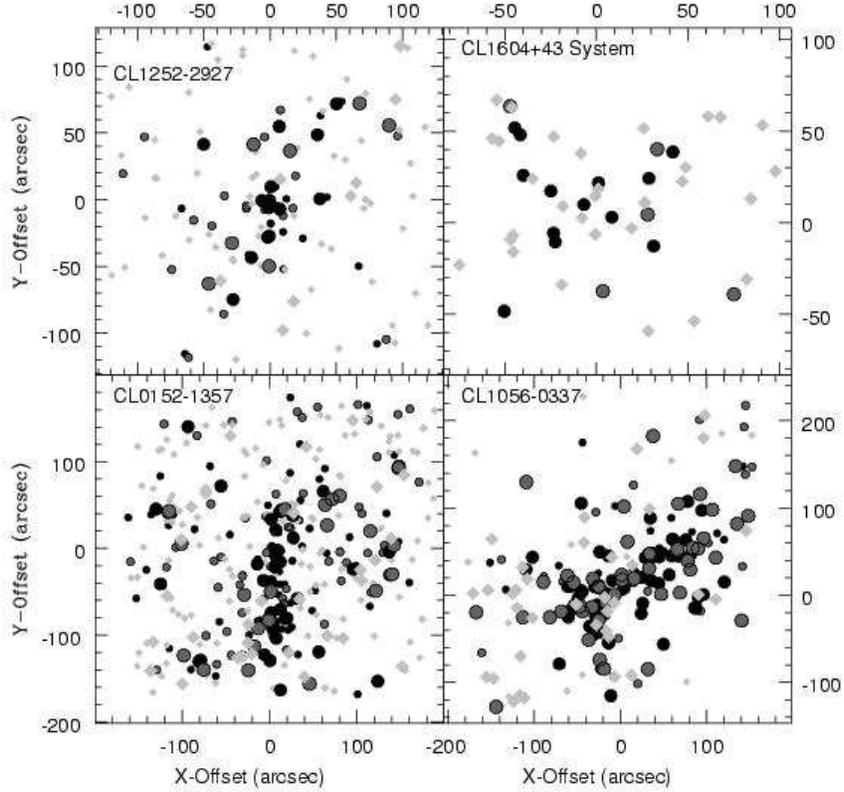}}
\caption{The spatial distribution of galaxy types in five
clusters. Light grey diamonds are spirals, dark grey circles are S0s,
and black circles are ellipticals.  Large symbols are
spectroscopically confirmed members and small symbols are candidate
cluster members based on Bayesian photometric redshifts.  Data for
the two 16-hr clusters are combined into one plot.}
\label{galtype_map}
\end{figure}

Figure \ref{galtype_map} shows the spatial distribution of
spectroscopically confirmed and photometrically selected galaxies in
CL1252, along with a number of other clusters in our sample, coded by
Hubble type.  The figure, which is a visual representation of the
morphology-density relationship (MDR), shows that E and S0 galaxies
are concentrated along an axis, whereas the latter type glaxies have a
much wider distribution.  The MDR for six clusters is discussed at
length in Section \ref{section_global}.

\begin{figure}[ht]
\centerline{\includegraphics[width=3.5in]{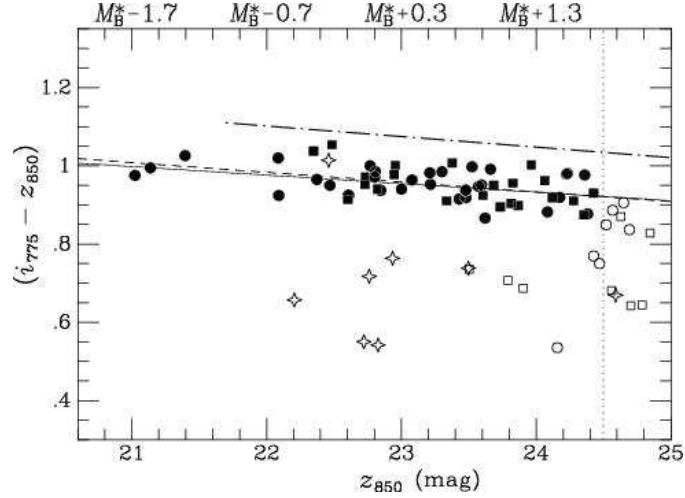}}
\caption{ACS color-magnitude diagram for confirmed members of the
CL1252 cluster, and other early-type galaxies within a 2 arcminute
radius of the cluster center (excluding spectroscopically known
interlopers).  Circles and squares represent elliptical and S0
galaxies, respectively. Solid symbols indicate galaxies that we use
for fitting the slope and scatter of the CM relation, while open
symbols (all of which lack spectroscopic information) were rejected as
probable interlopers or as below the faint magnitude cutoff (indicated
by the dotted line).  Finally, the star symbols show 8 confirmed
late-type members of the cluster, most of which are significantly
bluer than the early-type CM relation.  Two representative linear fits
are shown: a fit to the 15 confirmed elliptical members (solid line)
and to the 52 early-type red-sequence galaxies (including probable but
unconfirmed members).  The labels at top give the approximate
luminosity conversion, assuming the WMAP cosmology and $-1.4$ mag of
luminosity evolution, such that $M_B^* = -21.7$ (AB).  The relation
for the Coma cluster, transformed to these bandpasses at $z=1.24$ (no
evolution correction), is indicated by the dot-dashed line.}
\label{cl1252_cmr}
\end{figure}

Figure \ref{cl1252_cmr} shows the observed ACS F775-F850LP
color-magnitude (CM) relation for CL1252.  The relation is quite tight
and implies an intrinsic color scatter in these bandpasses of only
0.024 mag for the ellipticals, or 0.030 mag for all the early-type
galaxies (ellipticals and S0s).  Such a low scatter at this redshift
when the universe was less than 5 Gyr old implies either a very high
degree of synchronization in the formation of the stars in different
galaxies (an unlikely scenario given the stochastic nature of
hierarchical assembly), or that the galaxies are already advanced in
age so that the fractional age differences are small.  Given an
assumed form of the star formation history, it is possible to derive
the mean age of the galaxy population.

We simulated the evolution in the observed galaxy colors using two
simple star formation histories (Blakeslee \etal\ 2003: B03).  In the
first, the galaxies form in single bursts randomly distributed over
the interval between the epoch of recombination and some ending time
prior to the epoch at which the cluster is observed.  In the second,
the galaxies form stars at constant rates between randomly selected
times prior to the epoch at which they are observed.  The true star
formation history is likely somewhere between the extremes of these
single burst and constant formation models.  The two models imply mean
ages of 2.6 to 3.3 Gyr for the elliptical galaxies with a scatter in
age of about 35\%.  This means that the stars in these galaxies formed
over a period of about 1~Gyr, centered near a redshift of $z\approx3$.
The same models give a mean age of 1--2 Gyr for the S0 population,
with a scatter of $\sim50$\%, and of course the blue colors and large
scatter of the late-type galaxies imply ongoing star formation.  These
results are consistent with the features observed in the galaxy
spectra.  Thus, galaxies in protoclusters at $z\approx3$ must have
experienced very high rates of star formation, which declined sharply
to the modest levels observed near $z\sim1$.  ACS observations of
protocluster candidates at these high redshifts appear to be
consistent with this view (Miley \etal\ 2004).

Lidman \etal\ (2004) come to very similar conclusions from their
analysis of a CMD measured from deep VLT J,K$_s$ images of
CL1252. Using instantaneous single-burst solar-metallicity models,they
conclude that the average age of galaxies in the center of CL1252 is
2.7 Gyrs.


Rosati \etal\ (2004a: R04) combined Chandra and XMM-Newton observations of
CL1252 to measure the temperature, gas mass, metallicity, and
bolometric luminosity of CL1252 within a 60$^{\prime\prime}$ radius
(500 kpc).  Their results are summarized in Table
\ref{tab:cl1252}. The total mass in the last column is within a radius
of $536 \pm 40$ kpc.

\begin{table}[ht]
\caption[X-ray Properties of CL1252.]
{X-ray Properties of CL1252.}
\begin{tabular*}{\textwidth}{@{\extracolsep{\fill}}lccccc}
\sphline
\it $L_{[0.5-2.0]}$ &\it $L_{[Bol]}$ &\it $T_X$ &\it $Z_{gas}$  &\it $M_{gas}$ &\it $M_{tot}$\cr
\it 10$^{44}$ erg s$^{-1}$  &\it 10$^{44}$ erg s$^{-1}$ &\it keV &\it  $Z_{\odot}$ &\it 10$^{13}M _{\odot}$&\it 10$^{14}M _{\odot}$ \cr
\sphline
1.9$\sb{-0.3}\sp{+0.3}$ & 6.6$\sb{-1.1}\sp{+1.1}$  &	6.0$\sb{-0.5}\sp{+0.7}$ & 0.36$\sb{-0.10}\sp{+0.12}$ &	1.8$\sb{-0.3}\sp{+0.3}$ & 1.9$\sb{-0.3}\sp{+0.3}$\cr
\sphline
\end{tabular*}
\label{tab:cl1252}
\end{table}

\begin{figure}[ht]
\centerline{\includegraphics[width=4.5in]{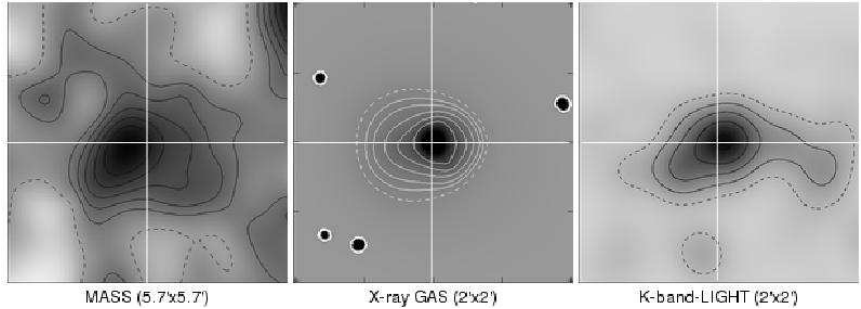}}
\caption{Mass, hot gas, and luminous matter in CL1252.  The left panel
is the projected mass distribution derived from L04's analysis of weak
lensing in ACS images. The middle panel is the X-ray contours from
adaptive smoothing of R04's Chadra observations of the cluster. The
right panel is T04's smoothed K-band light distribution of
photometrically selected cluster members. The images are rotated 90
degrees counter clock wise with respect to the images in Figures
\ref{cl1252_ctr} and \ref{cl1252_all}.}
\label{cl1252_x-ray}
\end{figure}

The left hand panel of Figure \ref{cl1252_x-ray} shows the projected
distribution of the total mass in CL1252 derived by Lombardi et
al. (2004:L04) from an analysis of the weak lensing in the ACS images.
The middle panel shows adaptively smoothed X-ray contours from R04's
Chandra observations of the cluster.  The right hand panel shows the
smoothed VLT K-band light distribution of photometrically selected
cluster members (Toft \etal\ 2004: T04).  The centroids
of the distributions of X-ray gas and galaxy light are very close to
one another. The adaptively smoothed X-ray image shows an edge
brightening on one side that suggests the gas and the brightest
concentration of galaxies are moving in a direction parallel to the
long axis of the cluster defined by its early-type galaxies (cf
Figures \ref{cl1252_ctr}, \ref{cl1252_all}, and \ref{galtype_map}). If
this interpretation is correct, the hot gas trapped in the deepest
potential is interacting with a lower density gas associated with
galaxies further down the axis of the cluster. However, the projected
distribution of total matter is lagging rather than leading the
compressed edge of the hot gas.  Unless the mass distribution is being
significantly affected by the mass associated with an obvious
foreground cluster, this fact argues against a scenario wherein
collisionless cold dark matter is leading hot gas that is retarded by
pressure forces.

Despite its age of less than 5 Gyr, R04 conclude that CL1252 is
well thermalized, with thermodynamical properties, as well as
metallicity, very similar to those of clusters of the same mass at low
redshift.  Nonetheless, the elongation and large angular extent of the
cluster, as well as the leading edge of the X-ray gas, suggest that it
is still collapsing, with mergers and gas stripping of many of the
galaxies yet in the future.  The relatively high value of the
metallicity is consistent with a scenario wherein the major episode of
metal enrichment and gas preheating by supernovae occurred at $z \sim 3$.

\subsection
{CL0152}

\begin{figure}[ht]
\centerline{\includegraphics[width=4.5in]{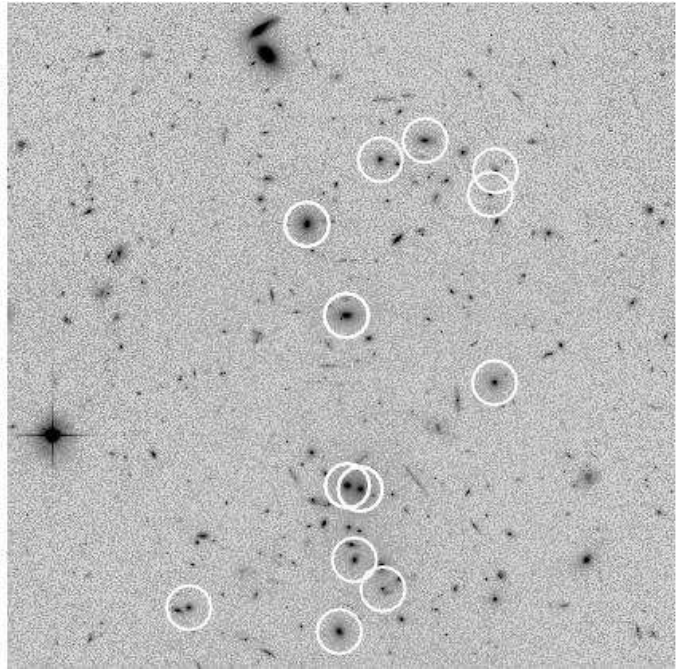}}
\caption{A composite ACS i,z image of the brighter of the two most
conspicuous sub-clusters in CL0152. The field size is
90$^{\prime\prime}$ ($\sim 690$ kpc in the restframe)
Spectroscopically confirmed members are circled (6$^{\prime\prime}$
diameter; $\sim 50$ kpc in the restframe).  There are several thin
arcs from lensed background galaxies that are much bluer than the
early-type galaxies in the cluster.  The lensed galaxy just below and
to the left of the two overlapping circles has two components that are
mirror images, indicating that the galaxy is very close to a caustic.}
\label{cl0152_ctr}
\end{figure}

\begin{figure}[ht]
\centerline{\includegraphics[width=4.5in]{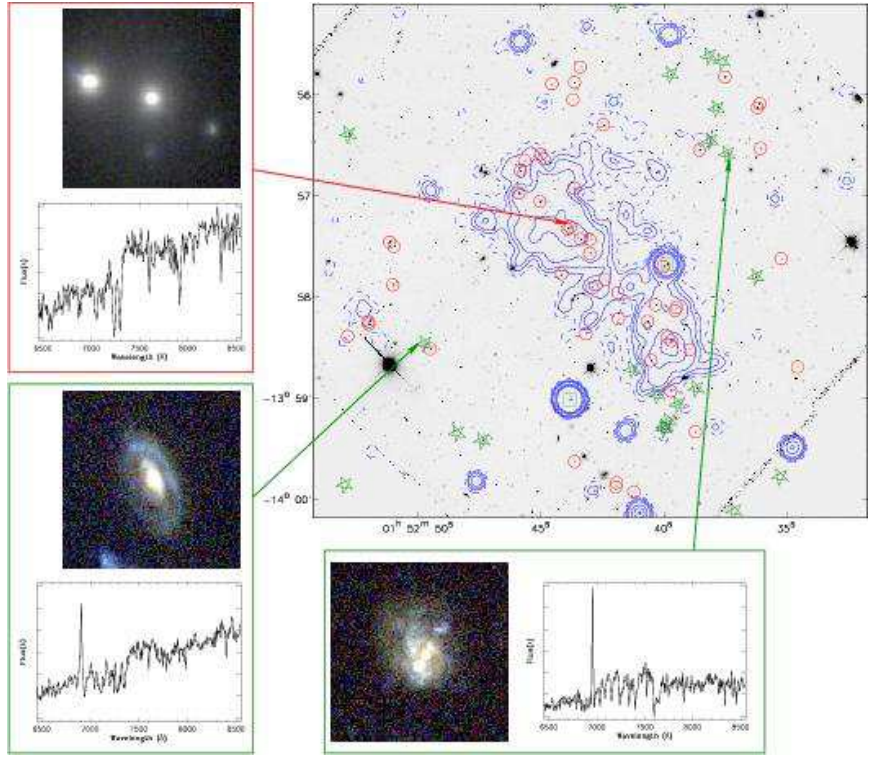}}
\caption{Star-forming galaxies in CL0152 and morphology of the X-ray
 emitting gas.  Spectroscopically confirmed passive galaxies are
 circled and galaxies with star formation, as indicated by
 [OII]$\lambda$3737 emission, are shown with a "star".  The latter are
 typically spirals and late-type galaxies, while the former are
 mostly E/S0s.  The insets show typical morphologies and spectra.
 The spatial segregation of the star forming later type galaxies is
 very striking.  The Chandra X-ray isophotes (Demarco \etal\ 2004;
 Maughan \etal\ 2003) are 3, 5, 7, 10, 20 and 30 sigma above the
 background.  }
\label{cl0152_starforming}
\end{figure}

\begin{figure}[ht]
\centerline{\includegraphics[width=4.5in]{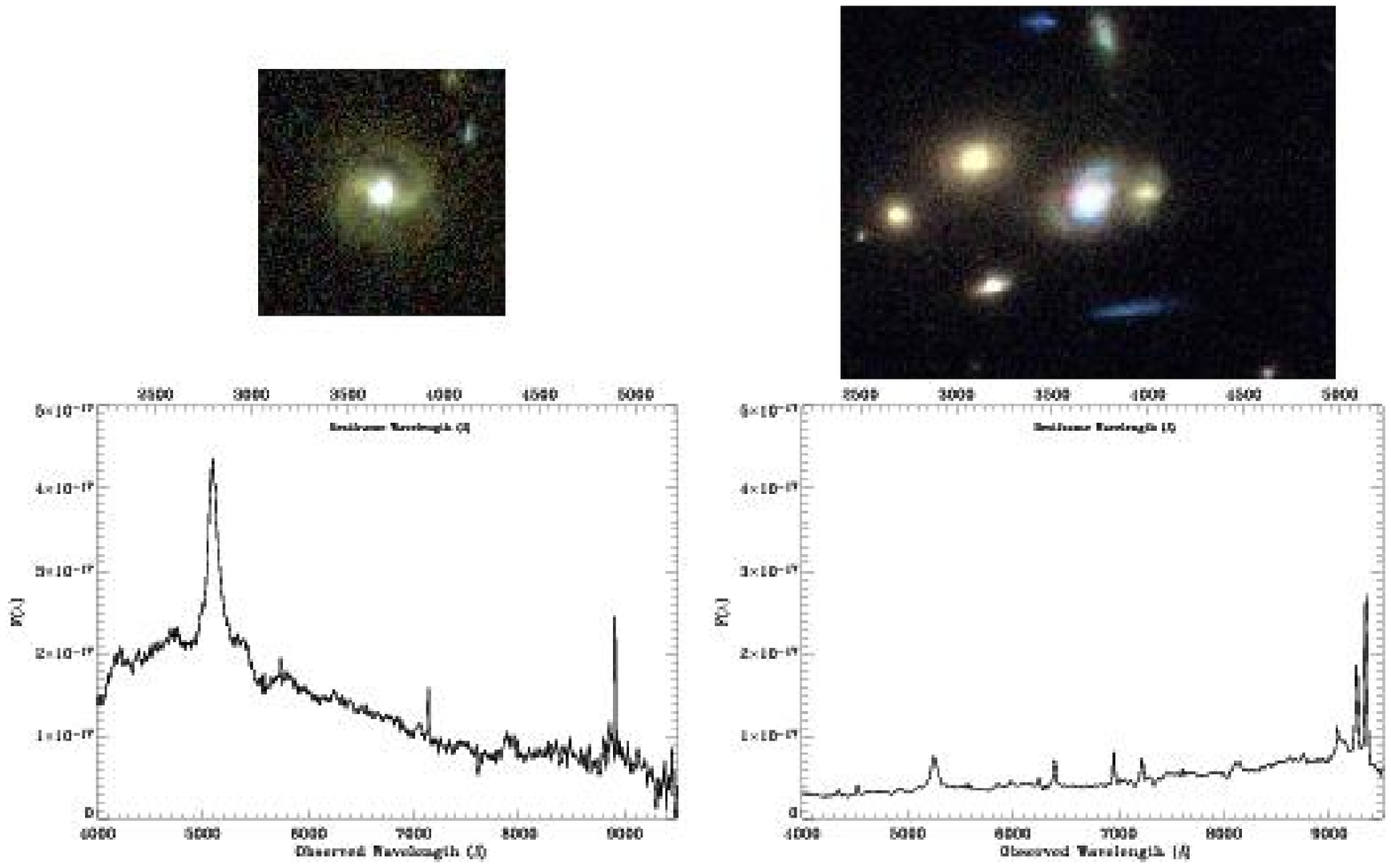}}
\caption{Two X-ray bright Seyfert galaxies in CL0152.  The respective
sizes of the left and right panels are $5^{\prime\prime} \times
5^{\prime\prime}$ ($38 \times 38$ kpc) and $10^{\prime\prime} \times
7.5^{\prime\prime} $ ($76 \times 57$ kpc).  The AGN in the right hand
panel is the brightest spiral and appears to be interacting with a
fainter, smaller spiral. The projected separation of the two spirals
is 1.15$^{\prime\prime}$ (8.8 kpc). The strong broad emission line in
each spectrum is MgII2800.}
\label{cl0152_agns}
\end{figure}

Figure \ref{cl0152_ctr} shows a composite ACS i,z image of the richest
of two prominent subclusters in CL0152. Spectroscopically confirmed
galaxies are circled.  Figure \ref{cl0152_starforming} shows an
overlay of Chandra X-ray contours on the entire ACS field. The
hot gas is confined to two components that coincide with two
concentrations of early-type galaxies.  As discussed below, the X-ray
properties and mass distribution derived from our weak lensing
analysis suggest that the two mass components are merging.  Confirmed
members with no emission lines are circled, and star-forming galaxies
with [OII]$\lambda$3727 emission are marked by "stars" (Homeier et
al. 2004).  The cluster is larger than the $\sim 350^{\prime\prime}$
($\sim2.7$ Mpc) field of the four overlapping ACS exposures.  The
confinement of the star-forming late-type galaxies to a ring or shell
around the two subclusters is very striking.  Two bright X-ray sources
in the field coincide with two galaxies that are confirmed members.
One of the two galaxies is an isolated (barred) spiral.  The other is
a very "disturbed" spiral in a compact group of five galaxies, and
appears to have had a recent close interaction with another galaxy.
Images and spectra of the two galaxies are shown in
\ref{cl0152_agns}. Both galaxies have broad MgII 2800 lines, showing
that they are Seyferts (Demarco \etal\ 2004).

Jee \etal\ (2004: J04) used ACS images of CL0152 to measure the
gravitation\-ally induced shear in the weekly distorted background
galaxies that fill the field around the cluster. The PSF of ACS has a
complicated shape, which also varies across the field. J04 constructed
the PSF model of ACS from an extensive investigation of 47 Tuc stars
in sufficiently uncrowded regions.  They verified that the model PSF
accurately describes the $actual$ PSF variation pattern in the cluster
observations after a slight adjustment of ellipticity is applied.

\begin{figure}[ht]
\centerline{\includegraphics[width=4.5in]{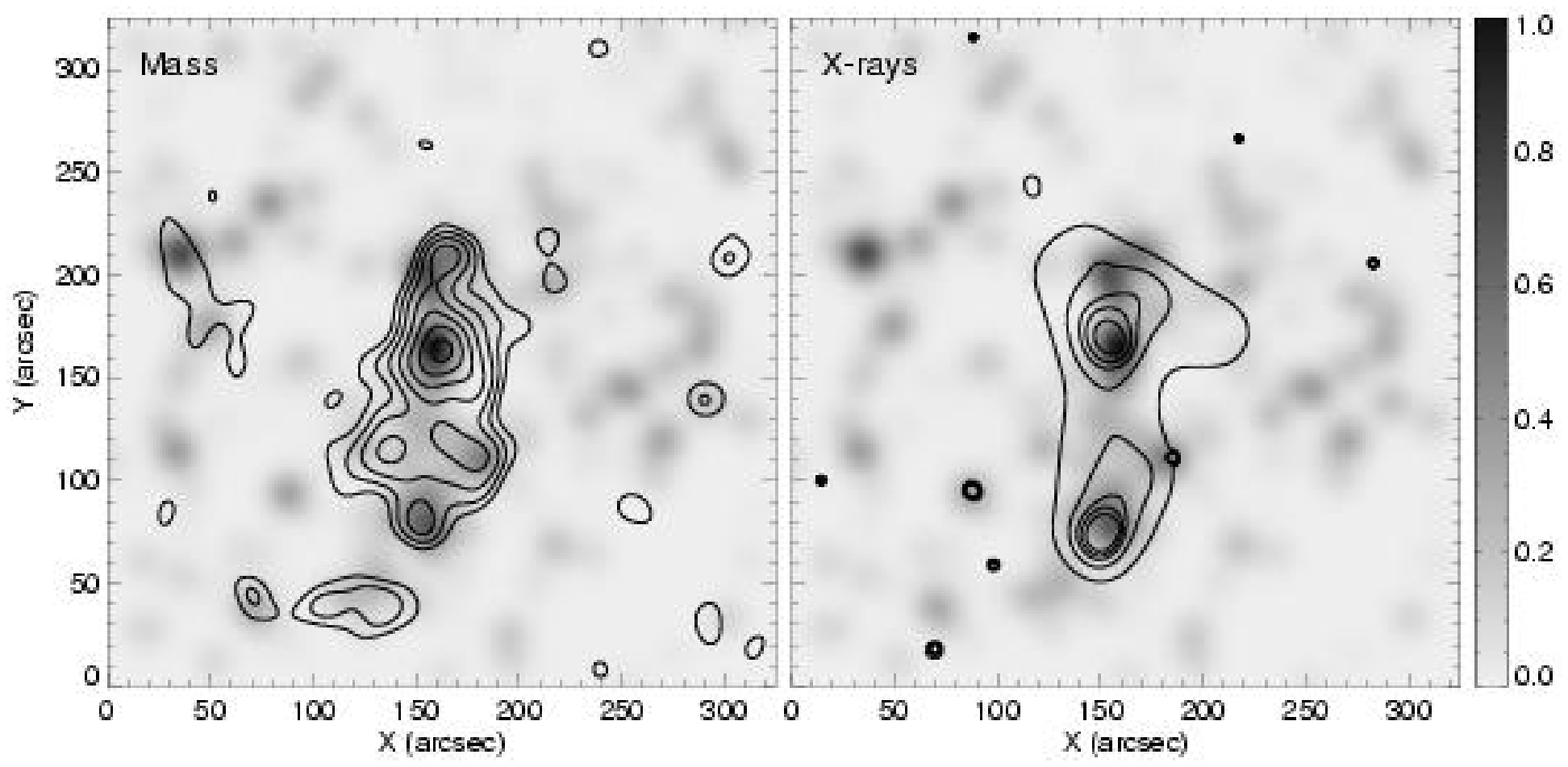}}
\caption{The left hand panel shows J04's mass contours overlaid on the
smoothed luminosity distribution derived from the $i_{775}$ image.
The map resolves five mass concentrations within the elongated main
body of the cluster.  There are an additional four mass concentrations
outside the main body that are associated with confirmed cluster
members. The right hand panel shows smoothed Chandra X-ray contours
overlaid on the luminosity distribution. The hot gas is lagging behind
the luminous and dark matter along and perpendicular to the axis of
the cluster.}
\label{mass_x-ray_lum}
\end{figure}

Figure \ref{mass_x-ray_lum}a shows the mass reconstruction created
from the shear field with a maximum-likelihood algorithm overlaid on a
smoothed luminosity distribution that is based on the
spectroscopically confirmed cluster members. The mass map is dominated
by the dark matter within the cluster. The correspondence between the
mass map and the luminosity distribution is quite good. Figure
\ref{mass_x-ray_lum}b shows the smoothed X-ray contours derived from
J04's reanalysis of archival Chandra observations. The figure shows
that the peaks of the X-ray emission in the two brightest subclusters
are lagging behind the peaks in the luminous matter and the dark
matter. The displacement of the southern X-ray peak relative to the
galaxies was noted by Maughan \etal\ (2003: M03).  They suggested that
the relatively collisionless galaxies are moving ahead of the gas. J04's
(dark) mass and luminosity maps strengthen this suggestion. The two
subclusters appear to be merging due to their mutual gravitational
attraction. If the mass is cold dark matter, it is collisionless,
whereas the merger of the hot gas in the two subclusters will be
slowed by pressure forces. Further support for this picture comes from
M03's suggestive evidence that there is a faint ridge of higher
temperature gas midway between the two components and perpendicular to
the long axis of the cluster.

In a standard $\Lambda$ CDM cosmology, J04 find a mass of $1.92\pm 0.3
\times 10^{14} M_{\odot}$ within a 50$^{\prime\prime}$ radius
(380kpc). This value agrees with M03's $2.4\sp{+0.4}\sb{-0.3} \times
10^{14} M_{\odot}$ within the same radius derived from the Chandra
X-ray observations. Transforming to $\Omega_M=0.3$,
$\Omega_{\Lambda}=0.7$, $H_0 =100$ for comparison with Joy \etal 's
(2001: J01) Sunyaev-Zeldovich mass of $2.1\pm0.7\times10^{14}
M_{\odot}$ within 65$^{\prime\prime}$, J04 find
$1.7\pm0.2\times10^{14} M_{\odot}$ for $r\le65^{\prime\prime}$, a
value encompassed by J01's larger error bars.

In summary, the relatively high spatial resolution of J04's weak
lensing mass distribution reveals several concentrations of mass that
coincide with luminous substructure within the cluster. The offsets
between the peaks in the X-ray emitting gas and the peaks in the dark
matter provide evidence that the cluster components are merging. The
mass derived from the weak lensing agrees with masses derived from the
SZ effect and from the X-ray emission.

\section
{Global Properties of the Clusters}
\label{section_global}

\noindent{\bf Cluster Morphology-Density Relation}
\par\smallskip\nobreak The correlation between morphology and density
is a fundamental characteristic of the local universe (Dressler 1980: D80; Postman \& Geller 1984: PG83; Goto et
al. 2003). The morphology-density relation (MDR) and its evolution are
therefore essential predictions of any viable large-scale structure
formation scenario.  Dressler \etal\ (1997: D97) measured
the MDR at $z \sim 0.4$ and concluded that the fraction of S0 galaxies
has increased significantly over the past 4 Gyr, suggesting that these
galaxies are relatively recent structures formed from later-type
systems infalling into clusters. Smith \etal\ (2004: Sm04)
used WFPC2 images of 6 clusters in the range $0.76 \le z \le 1.27$
(many of which are in common with our ACS survey) to obtain the first
measurement of the MDR at look back times up to 8.8 Gyr. Sm04 find
that the form of the MDR has undergone significant evolution
particularly at the high density end where the early-type (E+S0)
fraction has increased from $0.7\pm0.1$ at $\sim10^3$ galaxies
Mpc$^{-2}$ at $z \sim 1$ to $\ge 0.9$ at the present epoch. At low
densities, very little evolution is detected. They propose a series of
simple models to explain this trend and nearly all result in an
early-type population at $z \sim 1$ in which S0 galaxies are quite
rare (typically $\le 10$\% of the population).

We visually classified the morphologies of all galaxies in each of our
ACS images regardless  of position or color. The morphological
classification was performed on the full sample of $\sim\,$3,500
galaxies brighter than 24 mag by one of us (MP) but 3 other team
members classified a subset of 20\% of these to provide an estimate of
the uncertainty in the classifications.  The classifications were done
in the ACS band that samples at least part of the rest-frame $B$-band
in each cluster. Unanimous or majority agreement between all 4
classifiers in the overlap sample was achieved for 75\% of the objects
brighter than $i_{775} = 23.5$. There is no significant systematic
offset between the mean classification for the 3 independent
classifiers and the classification by MP giving confidence that the
full sample was classified in a consistent manner. MP also classified
all galaxies from our ACS exposure of MS1358 ($z=0.33$) that were in
common with the extensive study performed by Fabricant \etal\ (2000).
Agreement between the MP classifications and those from Fabricant et
al. was achieved $\sim 80$\% of the time with no systematic bias seen
in the discrepant classifications.

We compute a projected density using the same prescription as D80,
D97, and Sm04. For CL0152 and MS1054, we have a sufficient number of
spectroscopic redshifts that we can compute the MDR just using
confirmed members.  We correct the measured density for incompleteness
in our redshift survey and to match the same fiducial luminosity limit
used by D80 - although we allow for evolution of the characteristic
galaxy luminosity (e.g., Postman, Lubin, \& Oke 2001). For CL1252 at
$z=1.24$, we derive the MDR using a photo-z selected sample. We
confirm that the photo-z derived MDR is not significantly different
from the spectroscopic redshift result.  We have corrected the
densities for CL1252 for incompleteness and contamination due to the
fact the scatter in the photometric redshifts is significantly larger
than the cluster velocity dispersion. The MDR expressed as the 
early-type fraction (E+S0) as a function of local projected density is
shown in Figure 9. We also show the MDR's derived for the current epoch
(D80) and at $z \sim 1$ by Sm04. We find that our ACS-based MDR
exhibits less evolution than the Sm04 result but is still
significantly different from the current epoch MDR. The excellent
angular resolution of ACS allows us to make a direct measurement of
the S0 fraction.  We find S0 fractions of $0.31\pm0.10$ and
$0.19\pm0.20$ at $z = 0.83$ and $z = 1.24$, respectively. Both values
are higher than the extrapolations made by Smith et
al. based on the observed S0 fractions at $z=0.5$. However, the
difference is at best a 2 sigma result.  Furthermore, the ACS data
suggest a weak dependence of the S0 fraction on projected density that
is similar to what is seen today. These S0 fractions are comparable
with what D97 find at $z=0.4$ but are less than the local S0 fraction,
which is approaching 0.5\,--\,0.6 in cluster cores (D80, PG83, Poggianti
2001).  If these results hold up as we extend our analysis to the
entire cluster sample it would suggest an early ($z > 1.2$) formation
of a 20--30\% population fraction of lenticular systems that
undergoes a doubling in population only in the past 4 Gyrs. One caveat
is that the 3 clusters used in the analysis here are all relatively
X-ray luminous systems -- it remains to be seen if there is a
significant correlation between X-ray luminosity and the rate at which
the MDR evolves. A more thorough description of our MDR measurement
will be discussed by Postman \etal\ (2004).

\begin{figure}[ht]
\centerline{\includegraphics[width=4.0in]{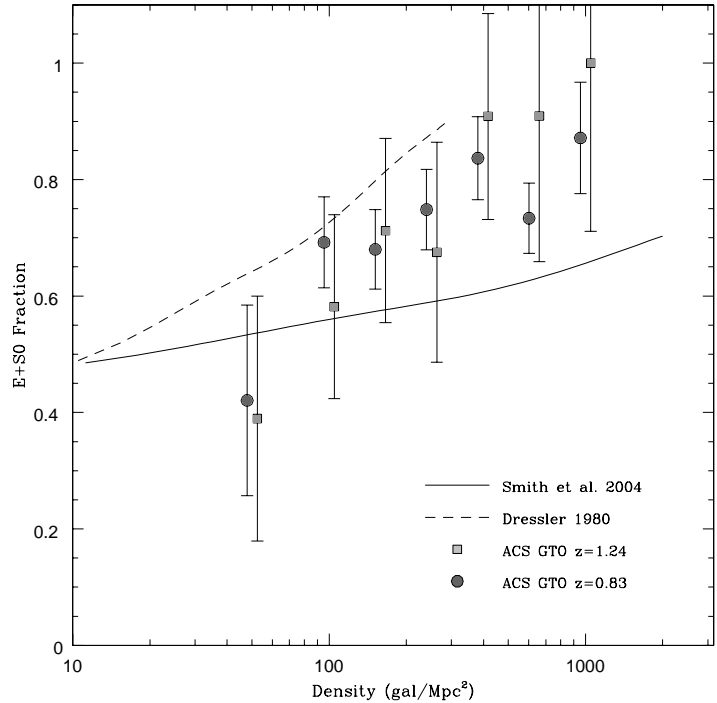}}
\caption{The morphology-density relation for two epochs: z=0.83 (7 Gyr
ago) and z=1.24 (8.6 Gyr ago). The early-type fraction includes all
galaxies classified as E or S0. Results from previous studies are
shown for comparison.  The Smith \etal\ (2004) result (based on WFPC2
imaging) covers a similar range of redshifts. We find a significantly
higher fraction of S0 galaxies at both epochs compared to their
extrapolation from z=0.50. We also see less evolution in the MDR than
they report but we are not inconsistent with their result.}
\label{MDR}
\end{figure}

\medskip
\noindent{\bf Brightest Cluster Galaxies}
\par\nobreak\smallskip\nobreak Studying the brightest cluster galaxies
(BCGs) in our clusters can reveal essential clues to the timescales
for their assembly process. The first significant difference is that 3
of the 6 BCGs are S0 or later. It is rare in current epoch clusters to
find such a high fraction of BCGs with disks. In the clusters with
later type BCGs, the process of galactic cannibalism may have either
not yet begun or may only just be getting underway.  We fitted the
2D surface brightness distributions of each BCG and used the best-fit 
model to measure the metric luminosity within a radius 
of 14.5 kpc ($h=0.7$, $\Omega_m=0.3$, $\Omega_{\Lambda}=0.7$). 
The photometric measurements were transformed to the
rest-frame AB B-band so that we could compare the data with
Bruzual-Charlot predictions for a passively evolving early-type
SED. We find that the BCGs in our clusters are, on average, less
luminous than a passively evolving BCG normalized to match the mean
current epoch BCG metric luminosity. The exception is the BCG for MS-
1054, which is quite consistent with these predictions. This suggests
that many of these BCGs will still undergo significant merger
events. For example, the residuals after our best fit model for the
BCG and 2nd-ranked in CL1252-29 are subtracted show substantial
structure consistent with tidally stripped stars. These two galaxies
are close together and thus appear to be in the process of merging. As
the 2nd-ranked galaxy is nearly as luminous as the BCG, the luminosity
of the final merger will be nearly double, making it comparable to the
luminosity of a current epoch BCG. A further indication that BCGs at 
$z \sim 1$ are still assembling is that the difference between the 1st
and 2nd-ranked luminosity within the above metric radius is on average
only 0.09 mag (again the MS-1054 M2-M1 value is the outlier at 0.30
mag).  Only 8\% of current epoch BCGs have M2-M1 as small as 0.09 mag
(Postman \& Lauer 1995).

\begin{figure}[ht]
\centerline{\includegraphics[width=4.5in]{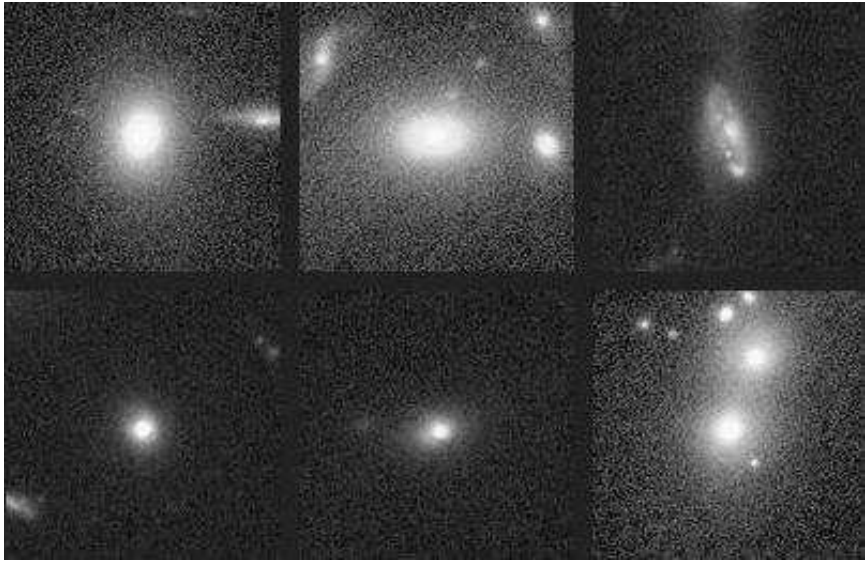}}
\caption{BCGs identified in 6 of our intermediate redshift
clusters. From left to right in the top row: CL0152-13, MS1054-03,
CL1604+4304. From left to right in the bottom row: CL1604+4321,
CL0910+54, CL1252-29.}
\label{bcg_mosaic}
\end{figure}

\begin{table}[ht]
\caption[Properties of the Brightest Cluster Galaxies.]
{Properties of the Brightest Cluster Galaxies in Six Intermediate Redshift Clusters.}
\begin{tabular*}{\textwidth}{@{\extracolsep{\fill}}lccccc}
\sphline
\it Cluster  &\it  RA(J2000)    &\it DEC(J2000)   &\it  $z_{obs}$  &\it  Type  &\it  Obs Mag \cr
 \it &\it Obs Color &\it  AB$B_{rest}$ &\it (U-B)$_{rest}$ &\it $(M2-M1)$ \cr
\sphline
MS1054    &   10:57:00.0 & -03:37:36 & 0.8313 & E  &  20.14 (i) \cr
 & 1.58 (V-i)&  -22.877   &  1.30   &  0.301 (i)\cr
CL0152     &  01:52:46.0 & -13:57:00 & 0.8342 & E  &  20.73 (i)\cr
 & 1.22 (r-i) & -22.516   &  1.26   &  0.088 (i)\cr
CL1604+4304 & 16:04:25.0 & +43:03:21 & 0.8966  & Sb/c & 20.89 (I)\cr
 & 1.65 (V-I) & -22.497  &   1.15  &   0.027 (I)\cr
CL1604+4321 & 16:04:36.7 & +43:21:41 & 0.9222 & S0/a & 21.31 (I)\cr
 & 1.15 (V-I) & -22.236  &   0.73  &   0.009 (I)\cr
CL0910   &    09:10:45.7 & +54:41:25  &   n/a &  Sa? & 21.52 (z)\cr
 & 1.03 (i-z) & -22.337   &  1.18   &  0.174 (z)\cr
CL1252   &    12:52:54.4 & -29:27:18 & 1.2343 & E   & 21.30 (z)\cr
 &0.96 (i-z) & -23.046  &   1.22  &   0.121 (z)\cr
\sphline
\end{tabular*}
\label{bcg_props}
\end{table}

\noindent {\bf Color-Magnitude Diagrams}
\par
\begin{figure}[ht]
\centerline{\includegraphics[width=4.5in]{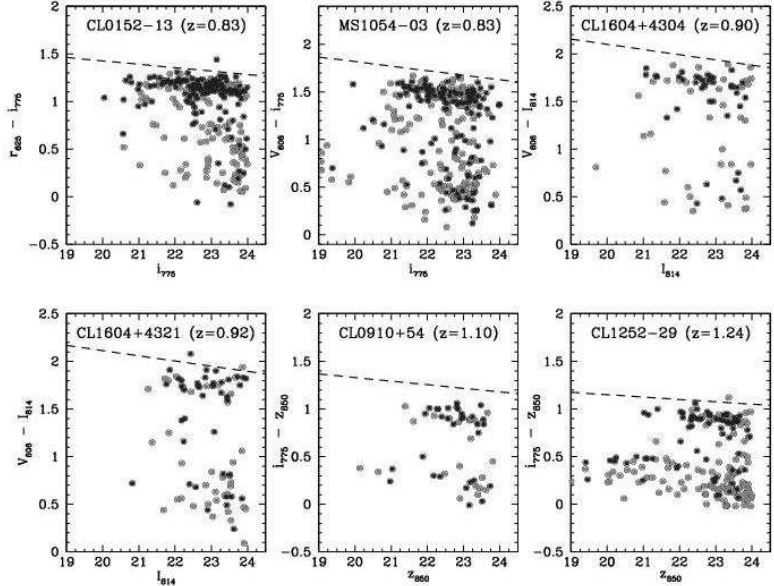}} 
\caption{The CMDs for E and S0 galaxies along the line of sight to six
$z \sim 1$ clusters.  Ellipticals are dark grey and the S0s are light
grey.  The dashed line is where the Coma Cluster CMD would lie if Coma
were simply redshifted to the relevant epoch and observed in the
indicated passbands.}
\label{cmd_all}
\end{figure}

\nobreak\smallskip\nobreak Figure \ref{cmd_all} shows that a well
defined "red sequence" appears to be a characteristic of all
clusters at $z \sim 1$.  Although we do not yet have final constraints
on the ages based on the scatter in color for all these systems,
the striking definition of the sequences suggests that, 
as with CL1252, we are observing the
ellipticals in these clusters at an epoch when they are 
$\sim 3$--4 Gyr old.

\section {Discussion and Summary}

Deep ACS and NICMOS images (this paper; van Dokkum \etal\ 2001) show
that elliptical, S0, and spiral galaxies are present in clusters at
redshifts up to $z \sim 1.3$. Analysis of the CMDs in clusters at $z
\sim 1$ (B03, L04, van Dokkum \& Stanford 2003, Holden \etal\ 2004)
suggest that the cluster ellipticals underwent the bulk of their
star formation 2.6 to 3.3 Gyrs earlier ($z \sim 3$).  The
morphology-density relation is well established at $z \sim 1$, with
star forming spirals segregated in the outer parts of clusters and
E/S0s concentrated in dense clumps.  Thus, the properties of the
majority of the luminous elliptical galaxies are well established.
The one exception is that the magnitude difference, $M2 - M1$, between
the second and first brightest galaxies in the clusters at $z \sim 1$
is smaller than in 93\% of present-day Abell clusters.

In contrast, the spatial distribution of early-type galaxies in the $z
\sim 1$ clusters is primarily elongated, often with two or more
dense concentrations of galaxies (sub-clusters and groups).
In general, the X-ray morphology follows the galaxy
concentrations, and, with the exception of MS1054, the X-ray emission
is not spherical (Rosati 2004b: R04b). In some clusters the X-ray
iso-contours suggest interactions between sub-clusters (e.g. CL0152).
There is evidence that the cluster bolometric X-ray luminosity $L_X$
evolves from high to low redshift, but there is little evolution in
the co-moving density (R04b). Finally, there appears to be mild
evolution of the $L_X$ vs $T_X$ relation (Ettori \etal\ 2004), in
disagreement with simple models.  The combination of the two results
suggests significant non-gravitational heating at earlier epochs.

When all of the facts are taken together, the following picture
emerges.  Clusters of galaxies are largely "in place" by $z \sim 1$,
an epoch when the Universe was $\sim\,$40\% its present age. The
early-type galaxies in these clusters, primarily found in high density
regions, had largely ceased star formation by the time the Universe
was 2.7 to 3.8 Gyrs old.  The intra\-cluster medium had already reached
the metallicity of present-day clusters by $z \sim 1$ (Tozzi \etal\ 2003).
The injection of metals from supernovae was likely one of the
mechanisms that heated the gas in addition to adiabatic compression as
the clusters collapse.  This injection of heat ``puffs up'' the gas,
which counteracts the expected cosmological evolution of the $L_X$ vs
$T_X$ relation, yielding the mild evolution observed.  The
distribution of the hot gas and early-type galaxies is very similar in
$z \sim 1$ clusters, suggesting that both have virialized in the
center of the deepest potentials in the clusters.  

In spite of the many similarities between $z \sim 1$ and present-day
clusters, there are many significant differences.  The distributions
of the hot gas and the early-type galaxies are irregular and 
elongated rather than spherical.  Thus, we appear to be observing the
clusters at an epoch when they were still assembling from groups
and sub-clusters.  Support for this picture
comes from the leading edge of the hot gas in CL1252 (R04) and from
CL0152 (J04) where the gas appears to be lagging behind the luminous
and dark mass in two merging sub-components.  The BCGs appear to have
considerable evolution ahead of them via mergers, for instance the
two central galaxies in CL1252 appear ready to merge in the near future.
Because spiral BCGs are rare in nearby clusters, the one in
CL1604+4304 is likely to undergo morphological transformation
or fade to lesser prominence as its gas is expended,
The overall merging process
will also result in the bolometric X-ray luminosities increasing as
the $z \sim 1$ clusters evolve. In spite of a strong selection bias
for the most luminous clusters, only one of our $z \sim 1$ clusters
(MS1054) is brighter than R04b's fiducial $L\sb{X}\sp{*} \sim 3 \times
10^{44}$ ergs sec$^{-1}$. Using R04b's value for the evolution in
luminosity, $L^{*} = L\sb{0}\sp{*}(1+z)^B$, with $B \sim -2.25$, the
clusters will brighten by approximately a factor of 5 by the present epoch.

Many tasks remain for the future.  We and others must fully
characterize all of the clusters that are being studied with ACS,
Chandra, XMM-Newton, and large ground based telescopes. Finally, we need to
find clusters at earlier stages of evolution between the present limit
$z \sim 1.3$ and the apparent proto-clusters being studied at
redshifts $z \ge 2$ (Miley \etal\ 2004).

\begin{acknowledgments}
ACS was developed under NASA contract NAS5-32865, and this research
was supported by NASA grant NAG5-7697.
\end{acknowledgments}








\end{document}